\documentclass[aps,prb,preprint,showpacs]{revtex4}
\usepackage{graphicx,longtable}

\begin{document}
\sloppy
\title{Superconductivity in Sr$_2$RuO$_4$ Mediated by Coulomb Scattering}
\author{Shigeru Koikegami}
\altaffiliation[Present address: ]{Nanoelectronics Research Institute, 
AIST Tsukuba Central 2, Tsukuba 305-8568, Japan}
\email[email address: ]{shigeru.koikegami@aist.go.jp }
\author{Yoshiyuki Yoshida} 
\affiliation{Japan Society for the Promotion of Science, 
6-Ichibancho, Chiyoda-ku, Tokyo 102-8471, Japan}
\author{Takashi Yanagisawa}
\affiliation{Nanoelectronics Research Institute, 
AIST Tsukuba Central 2, Tsukuba 305-8568, Japan}
\date{today}

\begin{abstract}
We investigate the superconductivity in Sr$_2$RuO$_4$ on the basis 
of the three-dimensional three-band Hubbard model. 
We propose a model with Coulomb interactions among 
the electrons on the nearest-neighbor Ru sites. 
In our model the intersite Coulomb repulsion and exchange coupling can work 
as the effective interaction for the spin-triplet paring.  This effective interaction 
is enhanced by the band hybridization, which is mediated by the interlayer transfers. We investigate the 
possibility of this mechanism in the ground state and 
find that the orbital dependent spin-triplet 
superconductivity is more stable than the spin-singlet one for 
realistic parameters. This spin-triplet superconducting state 
has horizontal line nodes on the Fermi surface. 
\end{abstract}


\maketitle
%
%

\section{Introduction}

The nature of the superconductivity in 
Sr$_2$RuO$_4$ has drawn much attention 
since its discovery in 1994.~\cite{Maeno1994,Maeno2001} 
A lot of experiments have provided evidence that the 
superconductivity is unconventional. For instance, the superconductivity 
is extremely sensitive to the non-magnetic impurity scattering in contrast 
to Anderson's theorem on a conventional superconductor.~\cite{Mackenzie1998} 
Miyake and Narikiyo have successfully shown that such an anomalous effect of impurity in 
Sr$_2$RuO$_4$ can be explained as an evidence of the spin-triplet pairing 
superconductivity.~\cite{Miyake1999} Nuclear magnetic resonance (NMR) measurement 
has revealed that the $^{17}$O Knight shift is almost unchanged 
in the transition into the superconducting phase.~\cite{Ishida1998} 
Furthermore, muon spin relaxation ($\mu$SR) time measurement~\cite{Luke1998} 
and polarized-neutron scattering study~\cite{Duffy2000} clarified that 
in the superconducting phase the time reversal symmetry is broken.    
From these experimental evidences it is almost confirmed that 
the superconductivity in Sr$_2$RuO$_4$ is the spin-triplet superconductivity. 
In the past few years, the momentum dependence of 
the superconducting gap function has become the central issue of this 
spin-triplet superconductor. In Sr$_2$RuO$_4$ 
the Fermi surface consists of three cylindrical pieces 
mainly originated from the four Ru-$4d$ electrons in 
three $t_{2g}$ orbitals.~\cite{Singh1995,Oguchi1995,Hase1996} 
Agterberg \textit{et al.} insisted that the temperature dependences of specific heat, 
penetration depth, and thermal conductivity 
can be explained by the orbital dependent superconductivity.~\cite{Agterberg1997} 
Additionally, recent specific-heat measurement at low temperature suggests the existence of line 
nodes.~\cite{NishiZaki2000}  

In order to determine the location of the nodes, we need the experimental 
results obtained by directional probes. In Sr$_2$RuO$_4$ 
the magnetothermal conductivity measurement seems 
the most powerful tool to investigate the location of the nodes.~\cite{Tanatar2001,Izawa2001} 
Two groups have reported that the thermal 
conductivity has no notable anisotropy when the magnetic field is applied to 
the direction parallel to the conducting plane. These results are quite different 
from the result of the cuprate superconductor, and they suggest that 
the pairing state with vertical line nodes has less possibility for the 
candidate in Sr$_2$RuO$_4$. Thus the paring state with horizontal line nodes 
seems to be appropriate to explain these experimental results for Sr$_2$RuO$_4$. 

Since Sr$_2$RuO$_4$ 
has single-layered perovskite structure as in the case for La$_{2-x}$Sr$_x$CuO$_4$, 
it has been supposed that its superconductivity is 
mediated by largely enhanced fluctuations common to these two-dimensional materials.
~\cite{Mazin1997,Monthoux1999,Arita1999,Nomura2000,Takimoto2000} 
However, it seems difficult to explain the spin-triplet 
paring state with horizontal line node. In order to solve this problem, 
Hasegawa \textit{et al.} listed the possible odd-parity states 
on the basis of the group-theoretical analysis.~\cite{Hasegawa2000} 
In their analysis they took notice of the body-centered-tetragonal lattice of Ru with 
lattice constants $a$ and $c$. And they insisted that in order to stabilize 
the gap function with the horizontal line node the effective interaction 
for electrons at $\mathbf{r}$ and ${\mathbf{r}}\pm (a/2)\hat{\mathbf{x}}
\pm (a/2)\hat{\mathbf{y}} \pm (c/2)\hat{\mathbf{z}}$ is crucial. 
Zhitomirsky and Rice have successfully shown that the gap function 
with the horizontal line node 
may lead to the temperature dependence of the specific heat observed 
in experiments.~\cite{Zhitomirsky2001} Futhermore, Annett \textit{et al.} 
have reproduced the experimental data of the superfluid density and the thermal conductivity 
on the basis of the multiband attractive Hubbard model with interlayer coupling.~\cite{Annett2002}

In this paper we propose that the superconductivity in Sr$_2$RuO$_4$ is mediated by 
Coulomb scatterings among the electrons at 
$\mathbf{r}$ and ${\mathbf{r}}\pm (a/2)\hat{\mathbf{x}}
\pm (a/2)\hat{\mathbf{y}} \pm (c/2)\hat{\mathbf{z}}$. 
Our model Hamiltonian is the three-dimensional (3D) 
three-band Hubbard model with quasi-two-dimensional character. 
Our microscopic description of the superconductivity in Sr$_2$RuO$_4$ may be 
considered as an application of the two-band mechanism superconductivity 
to the spin-triplet Cooper pairing,~\cite{JKondo1963} or as the three-dimensional 
version of the spin-triplet superconductivity in the one-dimensional chain with long-range 
attractive Coulomb interactions.~\cite{Aligia1999} 

%
%
\section{3D three-band Hubbard model}

We consider three $t_{2g}$ orbitals of Ru-$4d$ electron, i.e., $d_{xy}$, $d_{yz}$, and $d_{zx}$, 
in our 3D three-band Hubbard model. It is represented in real space as 
\begin{eqnarray}
\label{10}
H & = & \sum_{{\mathbf r}{\mathbf r}^\prime \varphi \varphi^\prime \sigma}\left[
\varepsilon_\varphi \delta_{{\mathbf r}{\mathbf r}^\prime} \delta_{\varphi \varphi^\prime}
-t_{\varphi \varphi^\prime}({\mathbf r},{\mathbf r}^\prime)\right]
c_{\varphi {\mathbf r} \sigma}^\dagger c_{\varphi^\prime {\mathbf r}^\prime \sigma} \nonumber \\
&   & +\sum_{{\mathbf r} {\mathbf r}^\prime \varphi \varphi^\prime \sigma \sigma^\prime}
U^{\sigma \sigma^\prime}_{\varphi \varphi^\prime}({\mathbf r},{\mathbf r}^\prime)
c_{\varphi {\mathbf r} \sigma}^\dagger
c_{\varphi^\prime {\mathbf r}^\prime \sigma^\prime}^\dagger
c_{\varphi^\prime {\mathbf r}^\prime \sigma^\prime} 
c_{\varphi {\mathbf r} \sigma} \nonumber \\
&   & +\sum_{{\mathbf r} {\mathbf r}^\prime \varphi \varphi^\prime \sigma \sigma^\prime}
J^{\sigma \sigma^\prime}_{\varphi \varphi^\prime}({\mathbf r},{\mathbf r}^\prime)
c_{\varphi {\mathbf r} \sigma}^\dagger
c_{\varphi^\prime {\mathbf r}^\prime \sigma^\prime}^\dagger
c_{\varphi {\mathbf r}^\prime \sigma^\prime} 
c_{\varphi^\prime {\mathbf r} \sigma} \nonumber \\
&   & +\sum_{{\mathbf r} {\mathbf r}^\prime \varphi \varphi^\prime \sigma \sigma^\prime}
K^{\sigma \sigma^\prime}_{\varphi \varphi^\prime}({\mathbf r},{\mathbf r}^\prime)
c_{\varphi {\mathbf r} \sigma}^\dagger
c_{\varphi {\mathbf r}^\prime \sigma^\prime}^\dagger
c_{\varphi^\prime {\mathbf r}^\prime \sigma^\prime} 
c_{\varphi^\prime {\mathbf r} \sigma},
\end{eqnarray}
where $c_{\varphi{\mathbf r}\sigma}$ 
($c_{\varphi{\mathbf r}\sigma}^{\dagger}$) 
is the annihilation (creation) operator of d electron with orbital 
$\varphi=\{xy,yz,zx\}$ and spin $\sigma=\{\uparrow,\downarrow\}$ on site ${\mathbf r}$. 
$\varepsilon_\varphi$ are site energies, as we set $\varepsilon_{zx\,(yz)}=\Delta>0$ and 
$\varepsilon_{xy}=0$.
$t_{\varphi \varphi^\prime}({\mathbf r},{\mathbf r}^\prime)$ are hopping integrals, 
as set 
\begin{eqnarray}
&   & t_{zx\, zx}({\mathbf r},{\mathbf r}\pm a{\mathbf {\hat{x}}})
=t_{yz\, yz}({\mathbf r},{\mathbf r}\pm a{\mathbf {\hat{y}}})=t_0, \\
&   & t_{zx\, zx}({\mathbf r},{\mathbf r}\pm a{\mathbf {\hat{y}}})
=t_{yz\, yz}({\mathbf r},{\mathbf r}\pm a{\mathbf {\hat{x}}})=t_1, \\
&   & 
t_{zx\, yz\,(yz\, zx)}({\mathbf r},{\mathbf r}\pm a{\mathbf {\hat{x}}}+a{\mathbf {\hat{y}}})  \nonumber \\
&   & \hspace{0.8em}=
-t_{zx\, yz\,(yz\, zx)}({\mathbf r},{\mathbf r}\pm a{\mathbf {\hat{x}}}-a{\mathbf {\hat{y}}}) = \pm t_2, \\
&   & t_{xy\, xy}({\mathbf r},{\mathbf r}\pm a{\mathbf {\hat{x}}})=
t_{xy\, xy}({\mathbf r},{\mathbf r}\pm a{\mathbf {\hat{y}}})=t_3, \\
&   & t_{xy\, xy}({\mathbf r},{\mathbf r}\pm a{\mathbf {\hat{x}}}\pm a{\mathbf {\hat{y}}})=t_4, \\
&   & t_{zx\, zx\,(yz\, yz)}\left({\mathbf r},{\mathbf r}\pm \frac{a}{2}{\mathbf {\hat{x}}}\pm 
\frac{a}{2}{\mathbf {\hat{y}}}\pm \frac{c}{2}{\mathbf {\hat{z}}}\right)=t_\perp^\prime,
\end{eqnarray}
and 
\begin{eqnarray}
&   & \hspace{-4em}t_{zx\,xy\,(xy\,zx)}\left({\mathbf r},{\mathbf r}\pm \frac{a}{2}{\mathbf {\hat{x}}}
+\frac{a}{2}{\mathbf {\hat{y}}}+ \frac{c}{2}{\mathbf {\hat{z}}}\right) \nonumber \\
& = & t_{zx\,xy\,(xy\,zx)}\left({\mathbf r},{\mathbf r}\pm \frac{a}{2}{\mathbf {\hat{x}}}
-\frac{a}{2}{\mathbf {\hat{y}}}- \frac{c}{2}{\mathbf {\hat{z}}}\right) \nonumber \\
& = & 
-t_{zx\,xy\,(xy\,zx)}\left({\mathbf r},{\mathbf r}\pm \frac{a}{2}{\mathbf {\hat{x}}}
+\frac{a}{2}{\mathbf {\hat{y}}}- \frac{c}{2}{\mathbf {\hat{z}}}\right) \nonumber \\
& = &  
-t_{zx\,xy\,(xy\,zx)}\left({\mathbf r},{\mathbf r}\pm \frac{a}{2}{\mathbf {\hat{x}}}
-\frac{a}{2}{\mathbf {\hat{y}}}+ \frac{c}{2}{\mathbf {\hat{z}}}\right) \nonumber \\
& = & 
t_{yz\,xy\,(xy\,yz)}\left({\mathbf r},{\mathbf r}+ \frac{a}{2}{\mathbf {\hat{x}}}
\pm \frac{a}{2}{\mathbf {\hat{y}}}+ \frac{c}{2}{\mathbf {\hat{z}}}\right) \nonumber \\
& = &  
t_{yz\,xy\,(xy\,yz)}\left({\mathbf r},{\mathbf r}- \frac{a}{2}{\mathbf {\hat{x}}}
\pm \frac{a}{2}{\mathbf {\hat{y}}}- \frac{c}{2}{\mathbf {\hat{z}}}\right) \nonumber \\
& = & 
-t_{yz\,xy\,(xy\,yz)}\left({\mathbf r},{\mathbf r}+ \frac{a}{2}{\mathbf {\hat{x}}}
\pm \frac{a}{2}{\mathbf {\hat{y}}}- \frac{c}{2}{\mathbf {\hat{z}}}\right) \nonumber \\
& = & 
-t_{yz\,xy\,(xy\,yz)}\left({\mathbf r},{\mathbf r}- \frac{a}{2}{\mathbf {\hat{x}}}
\pm \frac{a}{2}{\mathbf {\hat{y}}}+ \frac{c}{2}{\mathbf {\hat{z}}}\right) \nonumber \\
& = & t_\perp^{\prime \prime}.
\end{eqnarray}
Hereafter, we only consider the on-site interactions and the interactions among 
the nearest neighbors along the $c$ axis, because the interactions among the 
nearest neighbors on the conduction $ab$ plane are negligible due to screening. 
If we take $\{{\mathbf {\hat{r}}}_i\}_{i=1,\ldots,8}
=\{[\pm (a/2){\mathbf {\hat{x}}},\pm (a/2){\mathbf {\hat{y}}},\pm (c/2){\mathbf {\hat{z}}}]\}$, the Coulomb integrals in Eq.~(\ref{10}) turn out
\begin{eqnarray}
U^{\sigma \sigma^\prime}_{\varphi \varphi^\prime}({\mathbf r},{\mathbf r}^\prime) & = & U^0_{\varphi \varphi^\prime}(1-\delta_{\varphi \varphi^\prime}\delta_{\sigma \sigma^\prime})
\delta_{{\mathbf r}\,{\mathbf r}^\prime}+U^1_{\varphi \varphi^\prime}
\sum_{i=1}^8 \delta_{{\mathbf r}\,{\mathbf r}^\prime+{\mathbf {\hat{r}}}_i}, \\
J^{\sigma \sigma^\prime}_{\varphi \varphi^\prime}({\mathbf r},{\mathbf r}^\prime) & = & J^0_{\varphi \varphi^\prime}(1-\delta_{\varphi \varphi^\prime})
\delta_{{\mathbf r}\,{\mathbf r}^\prime}+J^1_{\varphi \varphi^\prime}
\sum_{i=1}^8 \delta_{{\mathbf r}\,{\mathbf r}^\prime+{\mathbf {\hat{r}}}_i}, \\
K^{\sigma \sigma^\prime}_{\varphi \varphi^\prime}({\mathbf r},{\mathbf r}^\prime) & = & \left(K^0_{\varphi \varphi^\prime}(1-\delta_{\varphi \varphi^\prime})
\delta_{{\mathbf r}\,{\mathbf r}^\prime}+K^1_{\varphi \varphi^\prime}
\sum_{i=1}^8 \delta_{{\mathbf r}\,{\mathbf r}^\prime+{\mathbf {\hat{r}}}_i}\right)
\delta_{\sigma -\sigma^\prime},
\end{eqnarray}
where $U_{\varphi \varphi^\prime}$, 
$J_{\varphi \varphi^\prime}$, and $K_{\varphi \varphi^\prime}$ are 
Coulomb repulsions, exchange interactions, and pair hoppings, respectively. 
Then we transform our Hamiltonian from the representation in real space into the one in momentum 
${\mathbf k}$ space by Fourier transform, and decompose it into $H=H_0+H^\prime$. 
The noninteracting part $H_0$ is represented by
\begin{widetext}
\begin{equation}
\label{2}
H_0=\sum_{{\mathbf k} \sigma}\left( 
c_{zx\,{\mathbf k}\sigma}^\dagger \,
c_{yz\,{\mathbf k}\sigma}^\dagger \,
c_{xy\,{\mathbf k}\sigma}^\dagger \right)\left(
\begin{array}{ccc} 
\varepsilon_{zx\,{\mathbf k}}+t_{\perp\,{\mathbf k}}^1  & t_{\parallel\,{\mathbf k}} & t_{\perp\,{\mathbf k}}^2 \\
t_{\parallel\,{\mathbf k}} & \varepsilon_{yz\,{\mathbf k}} +t_{\perp\,{\mathbf k}}^1 & t_{\perp\,{\mathbf k}}^3 \\
t_{\perp\,{\mathbf k}}^2 & t_{\perp\,{\mathbf k}}^3 & \varepsilon_{xy\,{\mathbf k}} \\
\end{array} \right)
\left( \begin{array}{c}
c_{zx\,{\mathbf k}\sigma} \\
c_{yz\,{\mathbf k}\sigma} \\
c_{xy\,{\mathbf k}\sigma} \\
\end{array} \right).
\end{equation}
\end{widetext}
In Eq.~(\ref{2}) we denote
\begin{eqnarray}
\varepsilon_{zx\,{\mathbf k}} & = & \Delta-2t_0\cos k_x-2t_1\cos k_y, \\ 
\varepsilon_{yz\,{\mathbf k}} & = & \Delta-2t_0\cos k_y-2t_1\cos k_x, \\ 
\varepsilon_{xy\,{\mathbf k}} & = & 
-2t_3( \cos k_x+\cos k_y)-4t_4 \cos k_x \cos k_y, \\
t_{\perp\,{\mathbf k}}^1 & = & -8t_\perp^\prime \cos \frac{k_x}{2} 
\cos \frac{k_y}{2} \cos \frac{k_zc}{2}, \\
t_{\perp\,{\mathbf k}}^2 & = & 8t_\perp^{\prime \prime}\cos \frac{k_x}{2}
\sin \frac{k_y}{2} \sin \frac{k_zc}{2}, \\
t_{\perp\,{\mathbf k}}^3 & = & 8t_\perp^{\prime \prime}\cos \frac{k_y}{2} 
\sin \frac{k_x}{2} \sin \frac{k_zc}{2}, 
\end{eqnarray} 
and $t_{\parallel\,{\mathbf k}} = -4t_2\sin k_x \sin k_y$,
taking the in-plane lattice constant as unity.
We can diagonalize $H_0$ with respect to the band indices $\zeta=\{\alpha,\beta,\gamma\}$ as 
$H_0=\sum_{{\mathbf k} \sigma}\sum_\zeta
\varepsilon_{\zeta\,{\mathbf k}}
a^\dagger_{\zeta\,{\mathbf k}\sigma}a_{\zeta\,{\mathbf k}\sigma}$
by orthogonal transformations, $c^\dagger_{\varphi{\mathbf k}\sigma}=\sum_\zeta
R_{\zeta \varphi {\mathbf k}}a^\dagger_{\zeta{\mathbf k}\sigma}$ and 
$c_{\varphi{\mathbf k}\sigma}=\sum_\zeta
R_{\varphi \zeta {\mathbf k}}a_{\zeta{\mathbf k}\sigma}$. 
The interacting part $H^\prime$ is represented by
\begin{widetext}
\begin{eqnarray}
H^\prime &  =  & 
\frac{1}{N} \sum_{{\mathbf k} {\mathbf k}^\prime {\mathbf q}} 
	\sum_{\sigma \sigma^\prime} \sum_{\varphi \varphi^\prime}
	U^0_{\varphi \varphi^\prime}
	(1-\delta_{\varphi \varphi^\prime}\delta_{\sigma \sigma^\prime})
    c_{\varphi {\mathbf k}+{\mathbf q} \sigma}^\dagger
	c_{\varphi^\prime {\mathbf k}^\prime-{\mathbf q}\sigma^\prime}^\dagger
	c_{\varphi^\prime {\mathbf k}^\prime \sigma^\prime} 
	c_{\varphi {\mathbf k} \sigma} \nonumber \\
&     & +\frac{1}{N} \sum_{{\mathbf k} {\mathbf k}^\prime {\mathbf q}} 
	\sum_{\sigma \sigma^\prime} \sum_{\varphi \varphi^\prime}
    J^0_{\varphi \varphi^\prime}
    (1-\delta_{\varphi \varphi^\prime})
    c_{\varphi {\mathbf k}+{\mathbf q} \sigma}^\dagger
	c_{\varphi^\prime {\mathbf k}^\prime-{\mathbf q}\sigma^\prime}^\dagger
	c_{\varphi {\mathbf k}^\prime \sigma^\prime} 
	c_{\varphi^\prime {\mathbf k} \sigma} \nonumber \\
&     & +\frac{1}{N} \sum_{{\mathbf k} {\mathbf k}^\prime {\mathbf q}} 
	\sum_{\sigma \sigma^\prime} \sum_{\varphi \varphi^\prime}
K^0_{\varphi \varphi^\prime}
(1-\delta_{\varphi \varphi^\prime})\delta_{\sigma -\sigma^\prime}
    c_{\varphi {\mathbf k}+{\mathbf q} \sigma}^\dagger
	c_{\varphi {\mathbf k}^\prime-{\mathbf q}\sigma^\prime}^\dagger
	c_{\varphi^\prime {\mathbf k}^\prime \sigma^\prime} 
	c_{\varphi^\prime {\mathbf k} \sigma} \nonumber \\
&     & +\frac{1}{N} \sum_{{\mathbf k} {\mathbf k}^\prime {\mathbf q}} 
	\sum_{\sigma \sigma^\prime} \sum_{\varphi \varphi^\prime}\left[
	U^1_{\varphi \varphi^\prime {\mathbf q}}
        c_{\varphi {\mathbf k}+{\mathbf q} \sigma}^\dagger
	c_{\varphi^\prime {\mathbf k}^\prime-{\mathbf q}\sigma^\prime}^\dagger
	c_{\varphi^\prime {\mathbf k}^\prime \sigma^\prime} 
	c_{\varphi {\mathbf k} \sigma}+
	J^1_{\varphi \varphi^\prime {\mathbf q}}
	c_{\varphi {\mathbf k}+{\mathbf q} \sigma}^\dagger
	c_{\varphi^\prime {\mathbf k}^\prime-{\mathbf q}\sigma^\prime}^\dagger
	c_{\varphi {\mathbf k}^\prime \sigma^\prime} 
	c_{\varphi^\prime {\mathbf k} \sigma}\right] \nonumber \\
&     & +\frac{1}{N} \sum_{{\mathbf k} {\mathbf k}^\prime {\mathbf q}} 
	\sum_{\sigma \sigma^\prime} \sum_{\varphi \varphi^\prime}K^1_{\varphi \varphi^\prime {\mathbf q}}
        \delta_{\sigma -\sigma^\prime}
        c_{\varphi {\mathbf k}+{\mathbf q} \sigma}^\dagger
	c_{\varphi {\mathbf k}^\prime-{\mathbf q}\sigma^\prime}^\dagger
	c_{\varphi^\prime {\mathbf k}^\prime \sigma^\prime} 
	c_{\varphi^\prime {\mathbf k} \sigma},
\end{eqnarray}
\end{widetext}
where $N$ is the number of ${\mathbf k}$-space points in the first
Brillouin zone (FBZ), and 
\begin{eqnarray} 
\label{1}
U^1_{\varphi \varphi^\prime {\mathbf q}}  & = & 8U^1_{\varphi \varphi^\prime}
\cos  \frac{q_x}{2} \cos  \frac{q_y}{2} \cos \frac{q_zc}{2}, \\
\label{7}
J^1_{\varphi \varphi^\prime {\mathbf q}}  & = & 8J^1_{\varphi \varphi^\prime}
\cos  \frac{q_x}{2} \cos  \frac{q_y}{2} \cos \frac{q_zc}{2}, \\
K^1_{\varphi \varphi^\prime {\mathbf q}}  & = & 8K^1_{\varphi \varphi^\prime}
\cos  \frac{q_x}{2} \cos  \frac{q_y}{2} \cos \frac{q_zc}{2}.
\end{eqnarray}

%
%
\section{Spin-triplet Superconductivity}

\begin{figure}
\includegraphics[width=6.5cm]{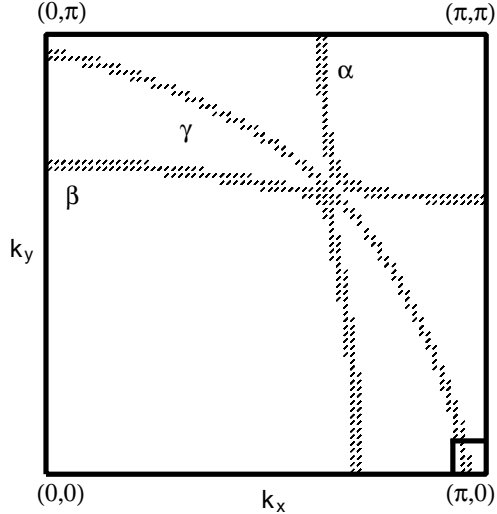}
\caption{\label{figure:1}Fermi surface in the case with $\Delta=0.50$. Band indices $\alpha$, 
$\beta$, and $\gamma$ are indicated here.}
\end{figure}

For our model we get a self-consistency equation for a gap function of the 
$\zeta$ band, $\Delta_{\zeta \mathbf{k}}$, 
within the weak-coupling formalism: 
\begin{equation}
\label{6}
\Delta_{\zeta \mathbf{k}} = 
-\frac{1}{2}\sum_{{\mathbf k}^\prime \zeta^\prime}
V_{\zeta \zeta^\prime {\mathbf k}{\mathbf k}^\prime}
\frac{\Delta_{\zeta^\prime {\mathbf k}^\prime}}%
{\sqrt{\left(\varepsilon_{\zeta^\prime {\mathbf k}^\prime}-\mu\right)^2
+\left|\Delta_{\zeta^\prime {\mathbf k}^\prime}\right|^2}}, 
\end{equation}
where $\mu$ is the chemical potential. Since our model does not include 
any asymmetrical interactions for spin state, e.g., spin-orbit interaction, 
this self-consistency equation is applicable 
to both spin-singlet and spin-triplet pairs in similar ways. 
For example, when we apply Eq.~(\ref{6})  to 
a spin-triplet pair taking its odd parity, i.e., 
$\Delta_{\zeta -\mathbf{k}}=-\Delta_{\zeta \mathbf{k}}$, 
into account, we get the expression of $V_{\zeta \zeta^\prime}$ as below:
\begin{eqnarray}
\label{3}
V_{\zeta \zeta^\prime {\mathbf k} {\mathbf k}^\prime} 
& = & \frac{2}{N}\sum_{\varphi \varphi^\prime}\left[
R_{\varphi \zeta {\mathbf k}}
R_{\varphi^\prime \zeta {\mathbf k}^\prime} U^1_{\varphi \varphi^\prime {\mathbf k}-{\mathbf k}^\prime}
R_{\zeta^\prime \varphi^\prime {\mathbf k}}
R_{\zeta^\prime \varphi {\mathbf k}^\prime} \right.\nonumber \\
&    & \left.\hspace{3em}+
R_{\varphi \zeta {\mathbf k}}
R_{\varphi^\prime \zeta {\mathbf k}^\prime}
J^1_{\varphi \varphi^\prime {\mathbf k}-{\mathbf k}^\prime}R_{\zeta^\prime \varphi {\mathbf k}}
R_{\zeta^\prime \varphi^\prime {\mathbf k}^\prime}\right].
\end{eqnarray}
On the other hand, in the case for a spin-singlet pair, $V_{\zeta \zeta^\prime}$ can be expressed as 
\begin{eqnarray}
\label{9}
V_{\zeta \zeta^\prime {\mathbf k}{\mathbf k}^\prime}
& = & \frac{2}{N}\sum_{\varphi \varphi^\prime}
\left\{R_{\varphi \zeta {\mathbf k}}
R_{\varphi^\prime \zeta {\mathbf k}^\prime} 
\left(U^0_{\varphi \varphi^\prime}
+U^1_{\varphi \varphi^\prime {\mathbf k}-{\mathbf k}^\prime}\right) R_{\zeta^\prime \varphi^\prime {\mathbf k}}
R_{\zeta^\prime \varphi {\mathbf k}^\prime} \right. \nonumber \\
&    & \hspace{3em}+R_{\varphi \zeta {\mathbf k}}
R_{\varphi^\prime \zeta {\mathbf k}^\prime}\left[
J^0_{\varphi \varphi^\prime}
(1-\delta_{\varphi \varphi^\prime})+J^1_{\varphi \varphi^\prime {\mathbf k}-{\mathbf k}^\prime}\right]
R_{\zeta^\prime \varphi {\mathbf k}}
R_{\zeta^\prime \varphi^\prime {\mathbf k}^\prime}
\nonumber \\
&    & \hspace{3em}\left.+
R_{\varphi \zeta {\mathbf k}}
R_{\varphi \zeta {\mathbf k}^\prime}\left[K^0_{\varphi \varphi^\prime}
(1-\delta_{\varphi \varphi^\prime})+K^1_{\varphi \varphi^\prime {\mathbf k}-{\mathbf k}^\prime}\right]
R_{\zeta^\prime \varphi^\prime {\mathbf k}}
R_{\zeta^\prime \varphi^\prime {\mathbf k}^\prime}\right\}.
\end{eqnarray}
When the gap magnitude $\Delta_{\mathrm{sc}}$ is small compared to band parameters, 
we can reduce Eq.~(\ref{6}) into 
\begin{equation}
\label{11}
\Delta_{\zeta \mathbf{k}} =
\ln \Delta_{\mathrm{sc}}\sum_{{\mathbf k}^\prime \zeta^\prime}
V_{\zeta \zeta^\prime {\mathbf k}{\mathbf k}^\prime}
\delta(\varepsilon_{\zeta^\prime {\mathbf k}^\prime}-\mu)
\Delta_{\zeta^\prime {\mathbf k}^\prime},
\end{equation}
according to the Kondo's argument.~\cite{JKondo2001} 
We choose our tight-binding band parameters as in Table~\ref{table:1}, where 
we take $t_0$ as a unit of energy estimated as about $1 {\mathrm{eV}}$. We choose them so that 
we can well reproduce the Fermi surface measured 
by the de Haas-van Alphen effect~\cite{Mackenzie1996,Yoshida1999,Bergemann2000} as shown 
in Fig.~\ref{figure:1}. Here we treat our tight-binding band parameters and Coulomb 
integrals as phenomenological ones. Thus it can be thought 
that our Fermi surface includes the band renormalization effects due to the electron correlation, 
and that the Coulomb integrals are effective interactions reduced by Hartree-Fock decoupling. 
Hartree-Fock decoupling also affects on-site energies, which we can control by varying $\Delta$. 
Our calculations are executed on equally spaced $256^3$ 
${\mathbf k}$ points in FBZ for each band. When we take $224^3$ 
${\mathbf k}$ points instead, our results of $\ln \Delta_{\rm sc}$ vary less than $3\%$.  
\begin{table*}
\begin{ruledtabular}
\begin{tabular}{ccccccccccccc}
$t_0$ & $t_1$ & $t_2$ & $t_3$ & $t_4$ & $t_\perp^\prime$ & 
$t_\perp^{\prime \prime}$ & $U^0_{\varphi \varphi^\prime}$ 
& $J^0_{\varphi \varphi^\prime}$ & $K^0_{\varphi \varphi^\prime}$ 
& $U^1_{\varphi \varphi^\prime}$ & $J^1_{\varphi \varphi^\prime}$
 & $K^1_{\varphi \varphi^\prime}$ \\ \hline
  $1.00$ & $0.12$ & $0.04$ & $1.00$ & $0.38$ & $0.01$ & $0.03$ &
 $2.00$ & $1.00$ & $1.00$ & $0.10$ & $0.10$ & $0.10$  \\
\end{tabular}
\end{ruledtabular}
\caption{\label{table:1}Transfers and Coulomb interactions.}
\end{table*}

When we solve our reduced self-consistency equation (\ref{11}), 
we find that the spin-triplet state is more stable than spin-singlet ones. 
One of this reason is that $U^1_{\varphi \varphi^\prime {\mathbf k}-{\mathbf k}^\prime}$ and 
$J^1_{\varphi \varphi^\prime {\mathbf k}-{\mathbf k}^\prime}$ in Eq.~(\ref{3}) 
can always change their signs due to their wave-vector dependences as shown in eqs.~(\ref{1}) and (\ref{7}). 
Added to this, the band hybridization enhances the effective interaction of Eq.~(\ref{3}) via 
the matrix elements of orthogonal transformations, $R_{\varphi \zeta {\mathbf k}}$ and 
$R_{\zeta \varphi {\mathbf k}}$, in Eq.~(\ref{3}). As a result for these, 
$U^1_{\varphi \varphi^\prime {\mathbf k}-{\mathbf k}^\prime}$ and 
$J^1_{\varphi \varphi^\prime {\mathbf k}-{\mathbf k}^\prime}$ work like strong pair tunneling interactions 
among hybridized bands for 
spin-triplet pairing. As shown afterward, this hybridization is much important for our 
triplet superconductivity. One of the most stable pairing functions is 
\begin{equation}
\label{8}
\Delta^x_{\zeta \mathbf{k}}=C_\zeta \, \sin \frac{k_x}{2} \cos \frac{k_y}{2} \cos \frac{k_zc}{2},
\end{equation}
where $C_\zeta$ is real and takes different value on each $\zeta$ band. Taking account of the spatial symmetry of our model, the other most stable function is 
\begin{equation}
\label{13}
\Delta^y_{\zeta \mathbf{k}}=C_\zeta \, \sin \frac{k_y}{2} \cos \frac{k_x}{2} \cos \frac{k_zc}{2}.
\end{equation}
It has indeed the same result of $\ln \Delta_{\rm sc}$ as the function, Eq.~(\ref{8}). 
These pairing functions, Eqs.~(\ref{8}) and (\ref{13}), have been 
proposed as candidates of the most stable state 
by Hasegawa {\textit{et al.}}~\cite{Hasegawa2000} 
In order to clarify the importance of the band hybridization, let us 
calculate the integrated effective matrix elements for our spin-triplet pairing function, Eq.~(\ref{8}),
\begin{equation}
v_{\zeta \zeta^\prime}=\sum_{{\mathbf k}}\,^\prime \hat{\Delta}^x_{\zeta {\mathbf k}} 
\sum_{{\mathbf k}^\prime} V_{\zeta \zeta^\prime {\mathbf k} {\mathbf k}^\prime}
\delta(\varepsilon_{\zeta^\prime {\mathbf k}^\prime}-\mu)
\hat{\Delta}^x_{\zeta^\prime {\mathbf k}^\prime},
\end{equation} 
where $\sum_{{\mathbf k}}\,^\prime$ denotes the momentum summation on the Fermi surface and 
$\hat{\Delta}^x_{\zeta {\mathbf k}}$ denotes the normalized function of (\ref{8}) determined by 
\begin{equation}
\sum_{{\mathbf k}}\,^\prime |\hat{\Delta}^x_{\zeta {\mathbf k}}|^2=1.
\end{equation}   
For example, when $\Delta=0.50$, we obtain 
\begin{equation}
\label{12}
\left(\begin{array}{ccc}
v_{\alpha \alpha} & v_{\alpha \beta} & v_{\alpha \gamma} \\
v_{\beta \alpha} & v_{\beta \beta} & v_{\beta \gamma} \\
v_{\gamma \alpha} & v_{\gamma \beta} & v_{\gamma \gamma} \\
\end{array} \right)=\left(
\begin{array}{ccc}
1.357 \times 10^{-3} & -1.175 \times 10^{-3} & -0.09952  \\ 
0.1057 & 2.093 \times 10^{-3} & 1.074 \times 10^{-3}  \\  
0.4129 \times 10^{-3} & 0.04082 & -2.015 \times 10^{-3} \\ 
\end{array} \right).
\end{equation}
Here, we can notice that the elements among the differnt bands $v_{\alpha \gamma}$, 
$v_{\gamma \beta}$, and $v_{\beta \alpha}$ have larger absolute values than the others. 
This is caused by the pair tunneling between the different bands, which is enhanced by the band hybridization. If we hope to 
increase our spin-triplet pairing instability, we should use these elements effectively. 
Judging from the inequalities, $v_{\alpha \gamma} < 0 < v_{\gamma \beta} < v_{\beta \alpha}$
, if $C_\alpha \cdot C_\beta < 0$ and $C_\beta \cdot C_\gamma < 0$ and $C_\gamma \cdot C_\alpha > 0$, 
we expect that the eigenvalue of Eq.~(\ref{11}), $(\ln \Delta_{\mathrm{sc}})^{-1}$, can take 
a large negative value for our gap function, Eq.~(\ref{8}). A large negative 
$(\ln \Delta_{\mathrm{sc}})^{-1}$ results large $\Delta_{\mathrm{sc}}$. Indeed, our numerically 
obtained solution of $C_\zeta$ shown in Table~\ref{table:2} satisfy the above inequalities. 
Hence the pair tunneling enhanced by the band hybridization 
plays a significant role to realize our spin-triplet superconductivity. 
\begin{table}
\begin{ruledtabular}
\begin{tabular}{cccc}
 $\zeta$ & $\alpha$ & $\beta$ 
& $\gamma$ \\ \hline 
$C_\zeta$ & $0.1625$ & $-0.1633$ & $0.08613$ \\ 
\end{tabular}
\end{ruledtabular}
\caption{\label{table:2}$C_\zeta$ of Eq.~(\ref{8}) in the case with $\Delta=0.50$.}
\end{table}

\begin{figure}
\includegraphics[width=7.7cm]{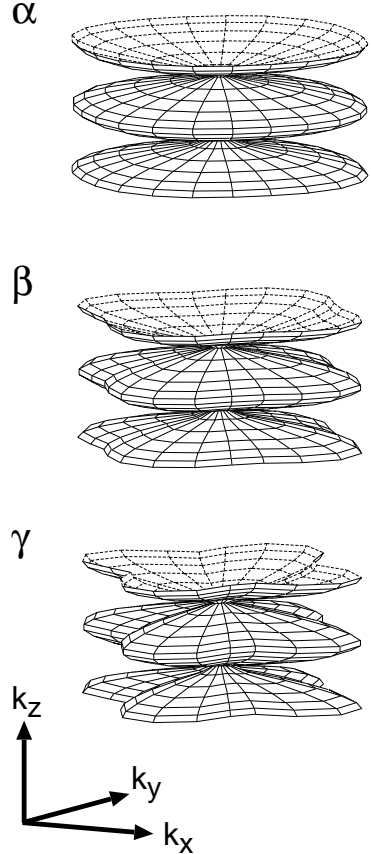}
\caption{\label{figure:2}Schematic pictures of gap amplitude on the Fermi surface 
of each band in the case with $\Delta=0.50$. The amplitude of each band is normalized in convenience.}
\end{figure}
Hereafter, we assume that 
the order parameter of spin-triplet superconductor with three components ($d$ vector) 
is parallel to the $z$ axis, ${\mathbf d}({\mathbf k}) \propto \hat{z}(k_x+{\mathrm i}k_y)$
.~\cite{Rice1995} Then, we can reasonably construct our $d$ vector as  
$d_z({\mathbf k}) = \Delta^x_{\zeta \mathbf{k}}+{\mathrm i} \Delta^y_{\zeta \mathbf{k}}$, which is a linear combination of our obtained functions, Eqs. (\ref{8}) and 
(\ref{13}).  We can show that the amplitude of $d$ vector vary as 
$\left|d_z({\mathbf{k}})\right| \propto 
\sqrt{1-\cos k_x \cos k_y}\left|\cos (k_zc/2) \right|$, 
shown in Fig.~\ref{figure:2}. All of them have holizontal line nodes at $k_z=\pm \pi/c$ and 
fourfold symmetries around the $c$ axis, and their amplitudes are larger along [100] and [010] 
than [110]. These results are qualitatively consistent with 
the magnetothermal conductivity measurements.~\cite{Tanatar2001,Izawa2001} 

\begin{figure}
\includegraphics[width=7.7cm]{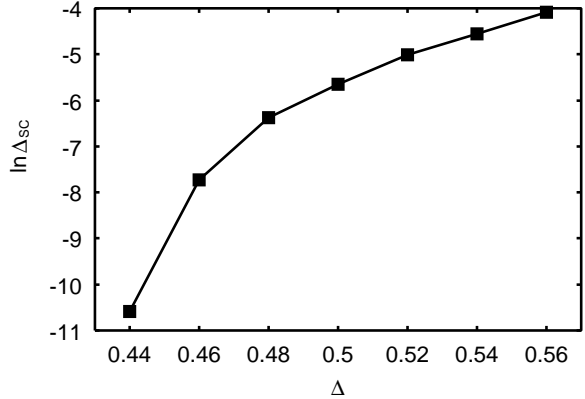}
\caption{\label{figure:3}$\Delta$-dependence of $\ln \Delta_{\mathrm{sc}}$.}
\end{figure}
\begin{figure}
\includegraphics[width=8.6cm]{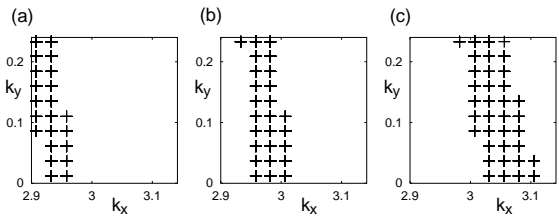}
\caption{\label{figure:4}Closeups of the Fermi surface projected 
on the plane with $k_z=0$. (a), (b), and (c) are in the cases with $\Delta=0.44$, $0.50$, and $0.56$, 
respectively. These areas are around the van Hove singular point as indicated in Fig.~\ref{figure:1}.}
\end{figure}
Then we study the $\Delta$-dependence of $\ln \Delta_{\mathrm{sc}}$. This result 
is shown in Fig.~\ref{figure:3}.  We show only the case with $0.44 \leq \Delta \leq 0.56$ 
because in other cases $\ln \Delta_{\mathrm{sc}}$ becomes extremely small. 
We can point out that our superconducivity is reinforced only when $\gamma$ band has a large density 
of states. To make 
this situation clear, we magnify the part of Fermi surfaces and project it on the plane with $k_z=0$. 
We show this part of all Fermi surfaces with different $\Delta$ in 
Fig.~\ref{figure:4}. The large density of states of the $\gamma$ band can be realized 
when a piece of Fermi surface is close to the van Hove singular point $(\pi,0)$. We 
have earlier shown that the pair tunneling enhanced by the band 
hybridization plays a significant role  for our spin-triplet superconductivity. Thus 
our spin-triplet superconductivity needs the two important factors.
It might be rare that both of these two factors present simultaneously in real materials. We can expect that 
in Sr$_2$RuO$_4$ both of these two conditions are wonderfully satisfied.   

In our results $\Delta_{\mathrm{sc}}$ can get to $e^{-4.084} \sim 16.8 {\mathrm{meV}}$. And, when a piece of the Fermi surface becomes closer to the van Hove singular point $(\pi,0)$, $\Delta_{\mathrm{sc}}$ will be much larger. These results are too much larger than the experimental results 
of Sr$_2$RuO$_4$, estimated as $0.2-0.4 {\mathrm{meV}}$. This may be caused by too large estimations of 
$U^1$ and $J^1$. However, we think that this is mainly caused by the weak-coupling formalism 
and neglected quasiparticles' lifetime. 
If the strong correlation effect decreases the lifetime, 
we should take into account the retardation effect and then $\Delta_{\mathrm{sc}}$ will be smaller. 
In Sr$_2$RuO$_4$ it is thought that the electrons correlate strongly with one another, 
and we should adopt the strong-coupling formalism for 
the quantitative estimation of $\Delta_{\mathrm{sc}}$.~\cite{Nomura2000} 
Although our quantitative estimation of $\Delta_{\mathrm{sc}}$ has these problems, 
as far as the whole electrons in Sr$_2$RuO$_4$ compose the Fermi liquid, 
our obtained gap symmetry cannot be replaced by the other symmetries. 

%
%
\section{Conclusion}

In this paper, we demonstrated that 
the spin-triplet pairing mediated by the intersite 
Coulomb scatterings is more stable than the 
spin-singlet one in our model. 
The gap function has a fourfold symmetry and 
horizontal line nodes on the Fermi surface of each bands. 
These results appear qualitatively consistent with the experimental 
results. Therefore the interlayer Coulomb scatterings play a significant role in order 
to realize the spin-triplet superconductivity in Sr$_2$RuO$_4$. Judged from 
the results about superconducting gap magnitude,  
our superconductivity is much sensitive to the band parameters.  
Our superconductivity is unique to  the electronic state in Sr$_2$RuO$_4$, which has 
both the degenerated orbitals and the interlayer transfers among these different orbitals. 

\begin{acknowledgments}
The authors are grateful to J. Kondo, 
K. Yamaji, M. Sigrist, K. Izawa, I. Hase, 
N. Shirakawa, S. I. Ikeda, and S. Koike for their invaluable comments. 
The computation in this work was performed on 
IBM RS/6000--SP at TACC and VT-Alpha servers at NeRI in AIST.
\end{acknowledgments}

\end{document}